# Machine learning algorithms to predict stroke in China based on causal inference of time series analysis

Qizhi Zheng[1#], M.D., Ayang Zhao[3#], Ph. D.; Xinzhu Wang[1], M.D., Yanhong Bai[1], B.S., Zikun Wang[1], M.D., Xiuying Wang[1], M.D., Xianzhang Zeng[2*], Ph. D.; Guanghui Dong[1*], Ph. D.

[1]College of Computer and Control Engineering, Northeast Forestry University, No.26, Hexing Road, Xiangfang District, Harbin, 150040, China.

[2]Department of Anesthesiology, Chongqing University Cancer Hospital, No.181, Hanyu Road, Shapingba District, Chongqing, 400030, China.

[3]School of Medicine and Health, Key Laboratory of Micro-systems and Micro-structures Manufacturing (Ministry of Education), Harbin Institute of Technology, Harbin, 150001 China

[#]These authors equally contributed to this work.

[*]**Correspondence author**
Name: Guanghui Dong
Mailing address: College of Computer and Control Engineering, Northeast Forestry University, No.26, Hexing Road, Xiangfang District, Harbin, 150040, China.
E-mail: dghRobert@126.com
Name: Xianzhang Zeng
Mailing address: Department of Anesthesiology, Chongqing University Cancer Hospital, No.181, Hanyu Road, Shapingba District, Chongqing, 400030, China.
E-mail: qwj0915@163.com

**Authors' contributions**

QZ and AZ conceived the study, participated in its design and coordination, and critically revised the manuscript. QZ and AZ had full access to all the data collection, analysis, and interpretation, and drafted the manuscript. XW, YB, ZW and XyW, contributed to the process of data collection and data analyses as study investigators. All authors approved the final manuscript. GD and XZ are the guarantors. All authors had full access to all the data in the study, and the corresponding authors had final responsibility for the decision to submit for publication. The corresponding author (GD and XZ) at tests that all listed authors meet authorship criteria and that no others meeting the criteria have been omitted.

**Abstract:**

**Importance:** Identifying and managing high-risk populations for stroke in a targeted manner is a key area of preventive healthcare.

**Objective:** To assess machine learning (ML) models and causal inference of time series analysis for predicting stroke clinically meaningful model.

**Design:** This is a retrospective cohort study and data is from China Health and Retirement Longitudinal Study (CHARLS) assessed 11,789 adults in China from 2011 to 2018. Data analysis was performed from June 1 to December 1, 2024.

**Setting:** CHARLS adopts a multi-stage probability sampling method, covering samples from 28 provinces, and collects data every two years through computer-aided personal interviews (CAPI).

**Participants:** This study employed a combination of Vector Autoregression (VAR) model and Graph Neural Networks (GNN) to systematically construct dynamic causal inference. Multiple classic classification algorithms were compared, including Random Forest, Logistic Regression, XGBoost, Support Vector Machine (SVM), K-Nearest Neighbor (KNN), Gradient Boosting, and Multi Layer Perceptron (MLP). The SMOTE algorithm was used to undersample a small number of samples and employed Stratified K-fold Cross Validation.

**Main Outcome(s) and Measure(s):** AUC (Area Under the Curve), Accuracy, Precision, Recall, F1 Score, and Matthews Correlation Coefficient (MCC).

**Results:** This study included a total of 11,789 participants, including 6,334 females (53.73%) and 5,455 males (46.27%), with an average age of 65 years. Introduction of dynamic causal inference features has significantly improved the performance of almost all models. The area under the ROC curve of each model ranged from 0.78 to 0.83, indicating significant difference ($P < 0.01$). Among all the models, the Gradient Boosting model demonstrated the highest performance and stability. Model explanation and feature importance analysis generated model interpretation that illustrated significant contributors associated with risks of stroke.

**Conclusions and Relevance:** This study proposes a stroke risk prediction method that combines dynamic causal inference with machine learning models, significantly improving prediction accuracy and revealing key health factors that affect stroke. The research results indicate that dynamic causal inference features have important value in predicting stroke risk, especially in capturing the impact of changes in health status over time on stroke risk. By further optimizing the model and introducing more variables, this study provides theoretical basis and practical guidance for future stroke prevention and intervention strategies.

**Trial Registration:** IRB00001052-11015.1.2

**Introduction**

Since 2015, stroke has become the leading cause of death and disability in China(1, 2). This poses a significant and growing threat to public health in China(3). Stroke is now a major chronic non-communicable disease requiring immediate and sustained attention. The Global Burden of Disease (GBD) Study 2019 reported 12.20 million incident cases, 101 million prevalent cases, and 6.55 million deaths from stroke worldwide(4). In China, these figures were 3.94 million, 28.76 million, and 2.19 million, respectively(5). Therefore, primary prevention of stroke has great public health importance.

Identifying and managing high-risk populations for stroke in a targeted manner is a key area of preventive healthcare(6, 7). This proactive approach allows for the implementation of specific interventions aimed at reducing the likelihood of stroke, significantly improving patient outcomes, and reducing the overall burden of the disease(8, 9). By combining risk factor assessment and diagnostic testing, high-risk individuals can be identified(10).

In recent years, machine learning (ML) has achieved remarkable advancements in the field of disease prediction, transforming the landscape of healthcare and offering promising avenues for improved diagnosis, treatment, and prevention(11-13). This progress stems from the increasing availability of large, high-quality datasets encompassing patient records, genetic information, medical images, and lifestyle factors. The ML models demonstrate the ability to predict the likelihood of developing various diseases, including cardiovascular disease(14), cancer(15), and diabetes(16), often with a considerable degree of accuracy. Early prediction allows for timely intervention, potentially preventing disease progression or mitigating its severity. For instance, ML algorithms can analyze electrocardiograms (ECGs) to detect subtle anomalies indicative of an impending heart attack, enabling prompt medical attention and potentially life-saving treatment(17). Similarly, in oncology, ML can analyze medical images to detect cancerous lesions at earlier, more treatable stages, improving patient outcomes significantly(18).

In this study, seven ML models, including random forest, logistic regression, XGBoost, SVM, KNN, Gradient Boosting and MLPKNN based on causal inference for time series analysis, were established to predict stroke in China.

**Research data and methods**

**Data sources**

The data used in this study is from China Health and Retirement Longitudinal Study (CHARLS), which has conducted a nationwide longitudinal follow-up survey on Chinese people aged 45 and above and their spouses since 2011. CHARLS adopts a multi-stage probability sampling method, covering samples from 28 provinces, and collects data every two years through computer-aided personal interviews (CAPI)(19). The data content includes health status, socio-economic status, lifestyle, etc., accompanied by physical measurements and blood sample collection. CHARLS provides rich longitudinal data resources, suitable for studying the dynamic trends of health and lifestyle changes in middle-aged and elderly populations.

This study used data from 2011 to 2018. The inclusion criteria are: ① Age ≥ 45 years old; ② Individuals with stroke related variables included in the data; The exclusion criteria are: ① Age<45 years old; ②Samples with missing values for any variable exceeding 10%. The CHARLS research protocol has been approved by the Biomedical Ethics Committee of Peking University (approval number: IRB00001052-11015.1.2).

**Handling the missing values**

To ensure data integrity and reliability of analysis results, this study adopted a systematic missing value processing strategy, as shown in Figure 1. Firstly, by calculating the proportion of missing values for each variable, variables with a missing ratio exceeding 10% are removed to reduce data bias. For the remaining missing values, a targeted filling method was adopted: for time series data, forward filling was used to ensure the temporal continuity of the data; For numerical variables such as age and weight, use mean imputation to preserve their statistical properties; For categorical variables such as gender and disease type, use mode padding and replace with the most common category. After completing the missing value processing, this study encoded the categorical variables. Specifically, for variables containing Chinese characters (such as provinces and cities), label encoding is used to convert them into numerical data for subsequent modeling and analysis. To address the issue of inconsistent scales of numerical variables, this study standardized them using Z-score, converting them into a standard normal distribution with a mean of 0 and a standard deviation of 1. This helps to eliminate scale differences between different variables, thereby improving the stability and effectiveness of model training.

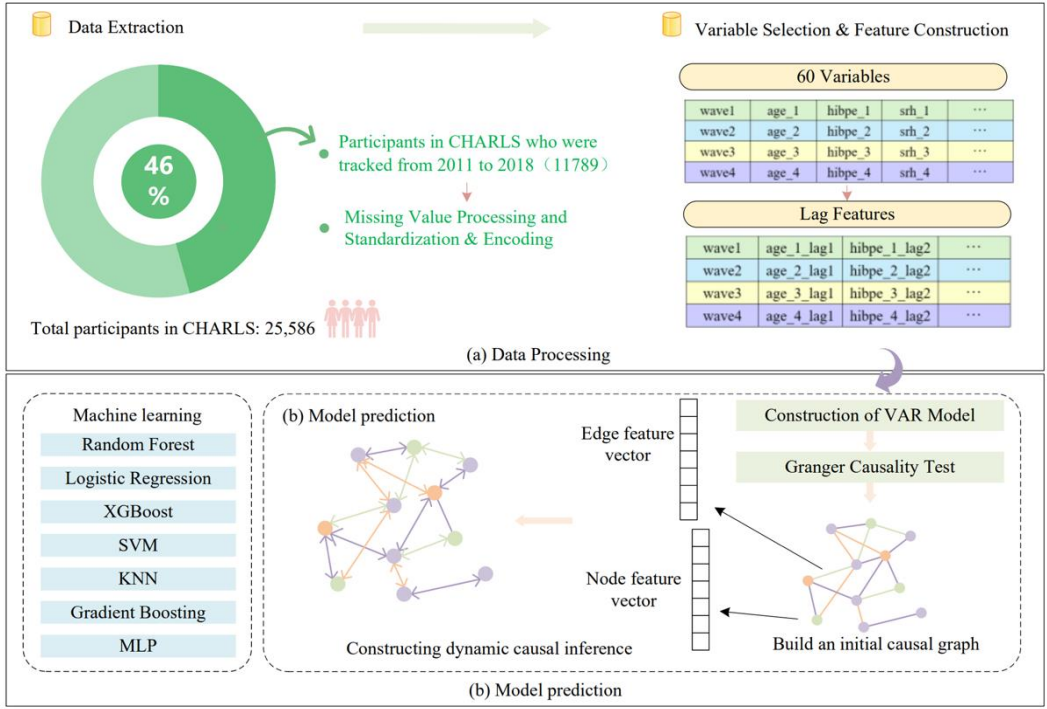

**Figure 1 (a) Data preprocessing and feature construction (b) Dynamic causal inference and model prediction**

**Construction of temporal features**

In order to comprehensively reveal the dynamic characteristics of health indicators and other related factors over time, this study introduced a time-series feature construction method in the data preprocessing stage. By constructing lagged and differential features, we can not only capture the historical impact of each variable, but also identify its changing trends.

The construction of Lag Features is a key step in time series analysis, aimed at introducing historical information to capture the dynamic changes of variables over time. In this study, we generated multiple lag features (e.g. lag 1, lag 2, etc.) for each health indicator that underwent feature selection. For the selected features $X^{(k)}$ (such as: age, BMI). The $l$-th lag feature $X^{(k)}_{lag_l}(t)$ is defined as the value of this feature at time point $t - l$, and formula (1) is as follows:

$$X^{(k)}_{lag_l}(t) = X^{(k)}(t - l) \tag{1}$$

$$X^{(k)}_{diff_l}(t) = X^{(k)}(t) - X^{(k)}_{lag_l}(t) \tag{2}$$

Among them, $t$ represents the current time point, and $l$ is the lag order (such as 1, 2). In addition to lagging features, we also constructed differential features to capture the trend of feature value changes. Differential feature $X^{(k)}_{diff_l}(t)$ is defined as the difference between the current value and its $l$-order lag value, as shown in formula (2). Differential features can reflect the rate of change of health indicators at different time points, which helps identify potential trends and patterns. In the dataset, each individual (identified by ID) is arranged in chronological order (identified by wave), generating multiple lag and differential features for each selected feature, thereby enriching the feature space and enhancing the model's ability to capture temporal dependencies.

**Dynamic causal inference**

In order to gain a deeper understanding of the dynamic dependency relationship between health indicators and other related factors and their impact on stroke occurrence, this study constructed dynamic causal inference. Dynamic causal inference not only reveals direct causal relationships between variables, but also combines lagged effects in time series and complex nonlinear relationships between variables. Causal analysis aims to identify causal effects between variables, not just correlations. For this purpose, this study employed a combination of Vector Auto-regression (VAR) model and Graph Neural Networks (GNN) to systematically construct dynamic causal inference.

**VAR model**

The VAR model models multivariate time series data and can simultaneously capture linear dynamic relationships between multiple variables, making it an effective tool for analyzing causal relationships in multivariate time series data. Firstly, we use the constructed lag features to establish a VAR model, specifically formula (3).

$$X_t = A_1 X_{t-1} + A_2 X_{t-2} + \cdots + A_p X_{t-p} + \varepsilon_t \tag{3}$$

$X_t$ is the vector of all variables at time point $t$, $A_1, A_2, \cdots A_p$ is the model coefficient matrix, $p$ is the lag order, and $\varepsilon_t$ is the error term. Then, the optimal lag order $p$ is selected through information criteria such as Akaike Information Criterion AIC and Bayesian Information Criterion BIC to ensure the model's goodness of fit and predictive ability. The process of selecting the lag order helps balance model complexity and prediction accuracy, avoiding overfitting or underfitting. Fit the VAR model based on the selected lag order and test the significance and stability of the model. This step ensures that the model can effectively capture the dynamic relationships between variables and lay the foundation for subsequent causal relationship testing.

After the VAR model fitting is completed, Granger causality test is performed to identify significant causal relationships. Granger causality test determines the existence of a causal relationship by examining whether the historical value of one variable can significantly improve the predictive ability of another variable. Assuming that for each pair of variables ($X_i$, $X_j$), we test whether $X_i$ is Granger causing $X_j$. Through this test, we identified causal relationships with p-values less than the predetermined significance level (such as 0.05), which will form the preliminary structure for dynamic causal inference.

**GNN model**

Although VAR models can reveal linear causal relationships between variables, in practical applications, there may be complex nonlinear relationships between variables. To this end, this study introduces Graph Neural Networks (GNNs) to further capture these complex relationships and optimize the construction of dynamic causal inference. To train the GNN model, it is necessary to construct a dataset that includes positive samples (significant Granger causality) and negative samples (variable pairs without causality). Negative samples are generated by randomly selecting variables that are not significant in the Granger test to ensure class balance in the training data and prevent the model from leaning towards predicting positive classes. We use the coefficients of the VAR model as node features to reflect the influence of each variable at different lag orders. If $A_k$ is the coefficient matrix of the first-order lag in the VAR model, then the eigenvector $f_i$ of each node can be expressed as:

$$f_i = [A_{1,i1}, A_{1,i2}, \cdots, A_{p,in}] \tag{4}$$

$A_{k,ij}$ represents the coefficient of influence of the *i*-th variable on the *j*-th variable in the *k*-th lag, and *n* is the total number of variables. To train the GNN model, we adopt the GCN architecture for predicting the existence of edges. GCN can effectively capture complex relationships between nodes in a graph by learning node representations through adjacency relationships and node features. The mathematical expression of the GCN model is:

$$H^{(l+1)} = \sigma\left(\tilde{D}^{-1/2} \tilde{A} \tilde{D}^{-1/2} H^{(l)} W^{(l)}\right) \tag{5}$$

$\tilde{A} = A + I$ is the adjacency matrix with self-loops added, $\tilde{D}$ is the degree matrix of $\tilde{A}$, $H^{(l)}$ is the node representation of the *l*-th layer, $W^{(l)}$ is the weight matrix of the *l*-th layer, and σ is the activation function. The model training aims to minimize the loss function of predicting edge existence, and this study adopts a binary cross entropy loss function.

Combine the analysis results of VAR model and GNN model to construct the final dynamic causal inference. Using the trained GNN model, predict all possible variable pairs and determine the existence of edges. Construct directed edges of variables predicted to have causal relationships, with nodes representing all relevant variables (including health indicators and other factors), thus forming dynamic causal inference. Dynamic causal inference not only reflects linear causal relationships between variables, but also integrates complex nonlinear relationships captured by GNN models.

**Machine Learning Modelling**

In order to comprehensively evaluate the performance of different machine learning models in stroke prediction, we compared multiple classic classification algorithms, including Random Forest, Logistic Regression, XGBoost, Support Vector Machine (SVM), K-Nearest Neighbor (KNN), Gradient Boosting, and Multi Layer Perceptron (MLP). These algorithms cover a variety of models from simple to complex, and can fully demonstrate the performance differences in stroke prediction tasks. To ensure the stability and reliability of the evaluation, we used the SMOTE algorithm to undersample a small number of samples and employed Stratified K-fold Cross Validation. This method ensures that the training and validation sets for each fold are consistent with the entire dataset in terms of class distribution, thereby avoiding potential biases caused by imbalanced data. In addition, we also group based on the hierarchical information of patient IDs to ensure that multiple data points of the same patient do not appear in different compromises, thereby avoiding potential data leakage issues and ensuring the independence and effectiveness of the evaluation.

**Statistical Analysis**

We used multiple evaluation metrics to comprehensively measure the performance of the model, including AUC (Area Under the Curve), Accuracy, Precision, Recall, F1 Score, and Matthews Correlation Coefficient (MCC). AUC reflects the overall performance of the model under different classification thresholds, accuracy measures the overall accuracy of classification, precision and recall focus on the model's ability to recognize positive and negative classes, respectively, while F1 Score and MCC comprehensively measure the model's balance and reliability.

**Results**

**Clinical Characteristics**

This study included a total of 11,789 participants, including 6,334 females (53.73%) and 5,455

males (46.27%), with an average age of 65.00 years. The average score of participants' Activities of Daily Living (ADL) was 0.51, and the average score of Instrumental Activities of Daily Living (IADL) was 0.64. The average score for self-assessment of health status (SRH) is 2.95 and most participants consider their health status to be good. Regarding disability status, 4,517 nondisabled participants (41.91%) and 5,228 disabled participants (44.35%) showed a certain proportion of elderly people facing disability issues. In terms of the prevalence of hypertension, 42.90% of the participants had hypertension. The prevalence of chronic diseases is relatively high, with 84.83% of participants having chronic diseases. In addition, smokers accounted for 74.10%, and the incidence of dyslipidemia was 76.58%.

**Table 1.  Clinical Characteristics**

| Variables | Total(n=11789) | non-stroke(n=10778) | stroke(n=1011) |
|---|---|---|---|
| Age, Mean±SD | 65.00±9.24 | 64.73±9.25 | 67.93±8.70 |
| Gender, n(%) | | | |
| female | 6334(53.73%) | 5814(53.94%) | 520(51.43%) |
| male | 5455(46.27%) | 4964(46.06%) | 491(48.57%) |
| ADL, Mean±SD | 0.51±1.19 | 0.42±1.05 | 1.45±1.97 |
| IADL, Mean±SD | 0.64±1.25 | 0.55±1.14 | 1.66±1.83 |
| srh, Mean±SD | 2.95±1.01 | 3.01±1.00 | 2.38±0.94 |
| disability, n(%) | | | |
| yes | 5228(44.35%) | 4517(41.91%) | 711(70.33%) |
| no | 6561(55.65%) | 6261(58.09%) | 300(29.67%) |
| hibpe, n(%) | | | |
| yes | 5058 (42.90%) | 4343 (40.30%) | 715 (70.72%) |
| no | 6731 (57.10%) | 6435 (59.70%) | 296 (29.28%) |
| chronic, n(%) | | | |
| yes | 10001 (84.83%) | 8993 (83.44%) | 1008 (99.70%) |
| no | 1788 (15.17%) | 1785 (16.56%) | 3 (0.30%) |
| smoken, n(%) | | | |
| yes | 8736 (74.10%) | 7927 (73.55%) | 809 (80.02%) |
| no | 3053 (25.90%) | 2851 (26.45%) | 202 (19.98%) |
| dyslipe, n(%) | | | |
| yes | 9028 (76.58%) | 8461 (78.50%) | 567 (56.08%) |
| no | 2761 (23.42%) | 2317 (21.50%) | 444 (43.92%) |
| memrye, n(%) | | | |
| yes | 11238 (95.33%) | 10403 (96.52%) | 835 (82.59%) |
| no | 551 (4.67%) | 375 (3.48%) | 176 (17.41%) |

**Dynamic causal inference and variable interpretation**

This study used a combination of Vector Autoregression (VAR) and Graph Neural Networks (GNN) to systematically construct a dynamic causal inference model. Through this method, we are able to capture the causal relationships between different variables over time and analyze in depth the impact of these characteristics on stroke risk. To construct a causal inference model, we first extracted basic features related to stroke risk from the data and extended these variables through lagged features. Through causal inference analysis of these characteristics, we have identified 10 key features directly related to stroke risk, which reflect multiple dimensions such as individual health status, daily living ability, and memory function. Figure 2 shows the causal relationships between these features, while Table 2 provides detailed descriptions of these features.

Figure 2 shows the causal relationships between the 10 key features selected through causal inference analysis. Each feature is represented as a node, and the arrows between nodes represent the causal flow between these features. Through this graphical display, we can intuitively observe which features play a direct role in predicting stroke risk. For example, self-rated health (SRH) directly affects the ability to manage memory disorders (memrye) and activities of daily living (ADL and IADL), which in turn affect the risk of stroke.

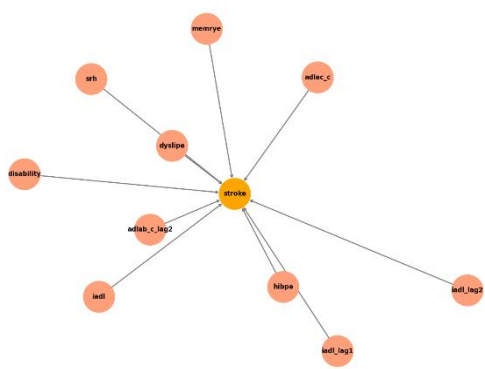

**Figure 2 Directly related features constructed by causal inference**

**Table 2. Basic Features Related to Stroke Risk**

| Features | Description |
| --- | --- |
| disability | Whether disabled |
| hibpe | Whether hypertensive |
| iadl | IADL (6 difficulties): housekeeping, cooking, shopping, making calls, taking medication, managing finances |
| srh | Self-rated health: excellent, very good, good, fair, poor, very poor |
| dyslipe | Whether dyslipidemia |
| memrye | Whether memory disease |
| adlab_c | ADL (6 difficulties): dressing, bathing, eating, getting in or out of bed, squatting and standing, and controlling urination and defecation |

### Predicting stroke using ML models

The Receiver Operating Characteristic Curve (ROC) of each model, visually presenting their performance under different classification thresholds. The larger the AUC value under the ROC curve, the stronger the model's ability to distinguish between positive and negative samples. From the graph, it can be seen that the introduction of dynamic causal inference features has improved the ROC curves of almost all models, showing higher AUC values, especially in the high threshold range (Figure 3).

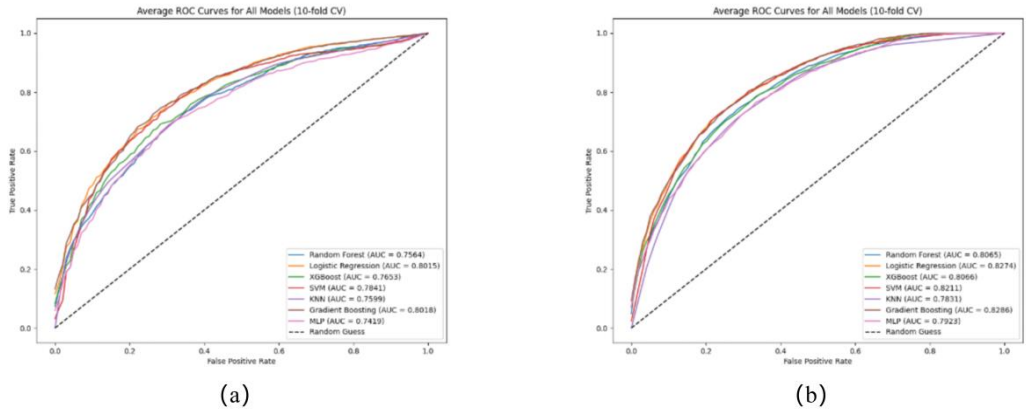

(a)  (b)

**Figure 3. Machine learning models for stroke prediction performance.** (a) ALL features;(b) Dynamic Causal Inference Features

Table 3 shows the evaluation results of each model under all features, and Table 4 shows the performance of the model after introducing dynamic causal inference features. It can be clearly seen from the comparison between Table 3 and Table 4 that the introduction of dynamic causal inference features has significantly improved the performance of almost all models. For example, the AUC of Logistic Regression under the original features is 0.8015, with an accuracy of 72.68%; After introducing dynamic causal inference features, the AUC increased to 0.8274 and the accuracy improved to 74.71%. This significant improvement indicates that dynamic causal features effectively enhance the model's ability to identify stroke risk, especially in distinguishing high-risk patients. Similarly, the AUC of Gradient Boosting increased from 0.8018 to 0.8286, and the accuracy increased from 72.75% to 74.34%, indicating that dynamic causal inference features play an important role in processing complex health-related data and capturing dynamic trends related to stroke. SVM and RF models have also significantly improved, with SVM's AUC increasing from 0.7841 to 0.8211 and accuracy increasing from 72.13% to 74.31%; The AUC of RF increased from 0.7564 to 0.8065, and the accuracy increased from 69.01% to 73.04%. These improvements indicate that the impact of dynamic causal features on these models is particularly important, especially in capturing changes in patient health status and predicting stroke risk. Although the performance improvement of the KNN model is relatively small, with AUC increasing from 0.7599 to 0.7831 and accuracy increasing from 68.63% to 71.61%, KNN still shows some improvement in processing simple data, indicating that the impact of dynamic causal features on different models varies.

**Table 3. Results of all feature evaluations**

| | ALL features | | | | | |
|---|---|---|---|---|---|---|
| Model | AUC | Accuracy | Precision | Recall | F1-Score | MCC |
| Random Forest | 0.7564 | 0.6901 | 0.6879 | 0.6987 | 0.6929 | 0.3807 |
| Logistic Regression | 0.8015 | 0.7268 | **0.7371** | 0.7062 | 0.7210 | 0.4544 |
| XGBoost | 0.7653 | 0.6960 | 0.7064 | 0.6719 | 0.6885 | 0.3926 |
| SVM | 0.7841 | 0.7213 | 0.7276 | 0.7076 | 0.7172 | 0.4431 |
| KNN | 0.7599 | 0.6863 | 0.7185 | 0.6164 | 0.6626 | 0.3773 |
| Gradient Boosting | **0.8018** | **0.7275** | 0.7352 | **0.7117** | **0.7230** | **0.4556** |
| MLP | 0.7419 | 0.6870 | 0.6977 | 0.6637 | 0.6793 | 0.3755 |

**Table 4. Evaluation results of introducing dynamic causal featuresDynamic Causal Inference Features.**

| Model | AUC | Accuracy | Precision | Recall | F1-Score | MCC |
|---|---|---|---|---|---|---|
| Random Forest | 0.8065 | 0.7304 | 0.7202 | 0.7552 | 0.7369 | 0.4617 |
| Logistic Regression | 0.8274 | **0.7471** | **0.7439** | 0.7552 | **0.7489** | **0.4950** |
| XGBoost | 0.8066 | 0.7233 | 0.7268 | 0.7169 | 0.7215 | 0.4469 |
| SVM | 0.8211 | 0.7431 | 0.7376 | 0.7562 | 0.7462 | 0.4873 |
| KNN | 0.7831 | 0.7161 | 0.7197 | 0.7085 | 0.7139 | 0.4325 |
| Gradient Boosting | **0.8286** | 0.7434 | 0.7365 | **0.7591** | 0.7472 | 0.4876 |
| MLP | 0.7923 | 0.7164 | 0.7228 | 0.7036 | 0.7128 | 0.4333 |

The dynamic causal inference feature reveals potential causal relationships between different health factors by analyzing the temporal changes in patients' health status. These features not only consider the health status of patients at various time points, but also track how these health changes affect the risk of stroke occurrence. Compared with static features, dynamic causal features provide richer information, enabling the model to capture long-term trends in health changes and improve the accuracy of stroke risk prediction.

Through the comparative analysis of the results in Tables 3 and 4, we conclude that the dynamic causal inference feature significantly improves the predictive ability of the model. Whether it is logistic regression, gradient boosting, support vector machine, or random forest, all models have improved in key indicators such as AUC, accuracy, and precision after introducing dynamic causal features. The dynamic causal inference feature helps the model better understand the relationship between changes in health status and stroke risk, providing a more accurate and reliable prediction tool for stroke prediction.

**Model Explanation and Feature Importance Analysis**

This study used the SHAP (Shapley Additive exPlans) method to measure the contribution of each feature to the model's predicted output. SHAP value is based on the concept of Shapley value, which considers each feature as a party in a "cooperative game" and measures its importance by evaluating the marginal contribution of the feature in all possible feature combinations. In this study, we ranked the importance of Gradient Boosting in predicting stroke by introducing dynamic causal features, as shown in Figure 4 (b). The results showed that disability, hibpe, iadl, and srh are the most important features for predicting stroke. Figure 4 (a) shows that features such as disability, hibpe, and iadl are positively correlated with stroke risk through SHAP values, indicating that functional impairment, uncontrolled blood pressure, and decreased daily self-care ability can all increase stroke risk. The positive development of lifestyle and psychological state (such as higher srh, increased frequency of social activity, muscle gain, and sufficient sleep) often shows a negative correlation with SHAP values, indicating that the increase of such positive factors can reduce the risk of stroke.

To illustrate the nonlinear relationship and key turning points between the changes in various feature values and the predicted results of the model (stroke risk). As shown in the main effect plot of IADL (Activities of Daily Living) in Figure 5, the red LOESS curve shows a continuous upward trend as the IADL score increases. When the iadl value is low, the SHAP value approaches 0, indicating that the impact on stroke risk is not significant at this time. But when the IADL score exceeds a certain critical point (indicated by the blue dashed line), the SHAP value rapidly increases, which means that when the independence of daily activities is further lost in elderly people, the risk prediction of stroke suddenly increases. This provides a warning for clinical and health services: strengthening rehabilitation support or the use of assistive devices when the IADL value approaches this critical point is expected to intervene early and delay the risk increase caused by functional decline.

Similarly, the hibpe features and their lagged values directly indicate whether an individual has hypertension. The SHAP main effect plot shows that when hibpe is 1 (i.e., having hypertension), the SHAP value is significantly positive, indicating that the presence of hypertension significantly increases the predictive probability of stroke risk. For the lagged values of hibpe, such as hibpe_1ag1 and hibpe_1ag2, similar trends were also shown, indicating that sustained hypertension over a continuous period of time has a cumulative positive impact on stroke risk. These findings emphasize the role of hypertension as an important modifiable factor in stroke risk and highlight the necessity of implementing effective blood pressure management strategies for hypertensive patients in clinical practice.

In summary, the value of these SHAP main effect maps lies in the intuitive presentation of the nonlinear impact mechanism of eigenvalues on stroke risk. Compared with global feature importance analysis and Summary Plot (Figure 4 (a)), these charts allow us to focus on a single feature and identify key change intervals at a fine numerical level, providing more accurate data support for the development of personalized prevention strategies and the timing of clinical interventions.

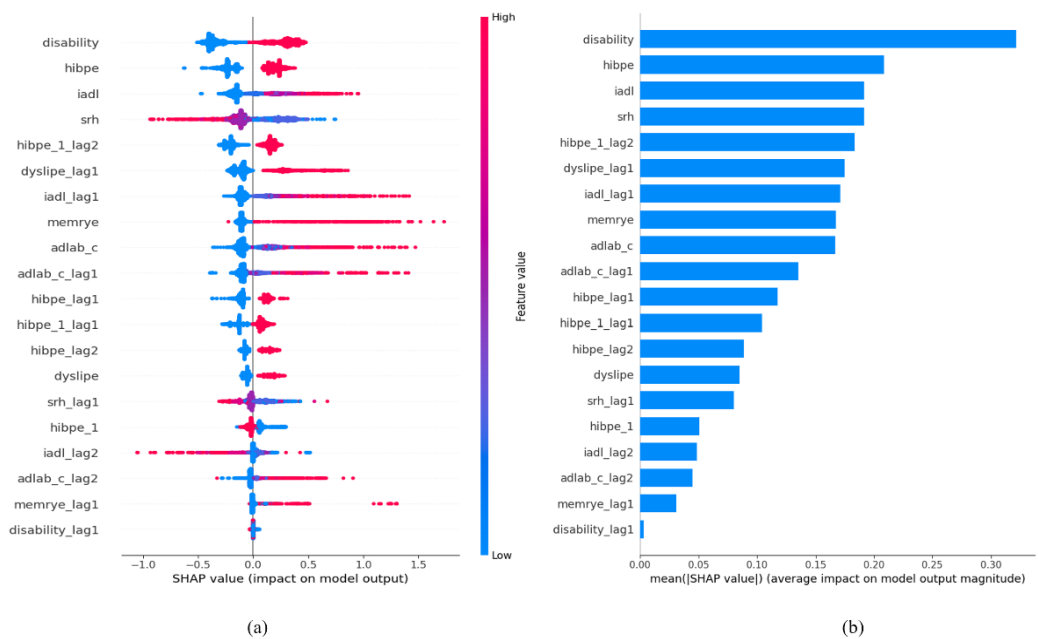

**Figure 4. Visual Explanation of Predictive Stroke Based on the Gradient Boosting Prediction Model.** (a)each dot represents an individual prediction, dot's position on the x-axis shows the impact that predictor has on the model's prediction for that individual. When multiple dots land at the same x position, they pile up to show density. The colour of the dot represents the level of the predictor related to that individual (colour reference on the right). (b)bar chart of average feature importance based on SHAP value magnitude.

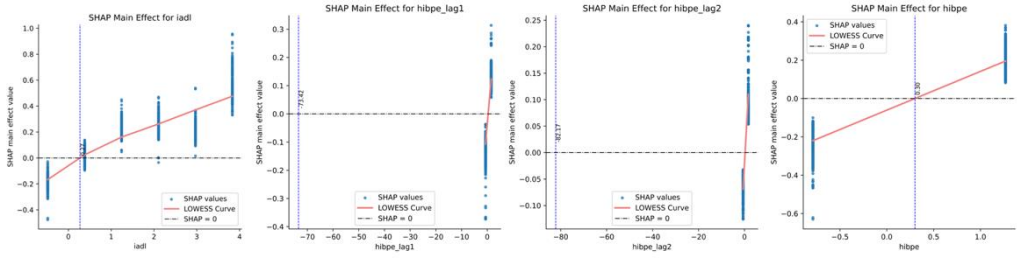

**Figure 5. SHAP main effect plot:** trend of the impact of eigenvalue changes on stroke risk. The horizontal axis represents the range of values for this feature, and the vertical axis represents the SHAP main effect value. The red LOWESS curve represents the trend of the relationship between changes in eigenvalues and SHAP values (the magnitude and direction of the contribution of this feature to the model's prediction of stroke risk). The vertical blue dashed line usually indicates the reference point or potential inflection point position of the feature value.

**Discussion**

This study proposes an innovative dynamic stroke risk prediction framework based on the CHARLS dataset, combined with causal inference and machine learning methods. The research results validated the key role of lagged health indicators in stroke risk prediction, particularly the potential impact of dynamic changes in long-term health factors such as IADL (Basic Activities of Daily Living) and blood pressure on stroke occurrence. By introducing dynamic causal inference features, the prediction accuracy of the model has been significantly improved, with an accuracy rate of 82.86%, demonstrating stronger predictive ability compared to traditional static models.

In the big data realm, AI techniques, such as machine learning, are revolutionizing the way physicians make clinical decisions and diagnosis, and have the potential to improve the estimated stroke risk scores to automate prediction(20). In traditional statistics, such as the logistic regression model, the c-statistic is a standard measure of predictive accuracy. However, machine learning, have several advantages for data analysis in predicting disease. One major benefit of most machine learning models is that no strict assumptions about the data distribution (e.g., normal distribution) are made. Within this context, most machine learning models can easily combine multi-omics data, including binary, categorial, discrete, and continuous variables without the need of extensive data preprocessing. Due to the regularization used in many machine learning methods, most machine learning models can also handle noisy data and large variances within the dataset comparatively well, although of course less noise is always preferred. Another benefit of machine learning models is that there are specialized types and architectures that can be trained on small datasets, especially those for which the number of features considerably outnumber the number of observations. At the same time, complex machine learning models can identify multi-faceted, non-linear patterns in the training data, which might not be obvious to human observers or simple linear models(21).

This study further revealed the dynamic impact of lagged health indicators such as age and BMI on stroke risk, indicating that historical changes in health status can significantly predict future stroke risk. Lag features not only consider the current health status, but also capture the trend changes in health status. This discovery has significant implications for dynamic health management strategies, as it can help clinicians identify individuals at high risk due to changes in their health status and provide scientific evidence for timely intervention.

By introducing dynamic causal inference features, we have successfully constructed an efficient stroke risk prediction model, providing solid support for early warning and intervention. With the arrival of an aging society, the incidence of stroke is increasing year by year. Early prediction and intervention are of great significance in reducing the incidence of stroke. This study provides clinical doctors with more precise tools to dynamically adjust treatment plans based on the patient's health status, thereby effectively reducing the incidence of stroke.

The order of lag features is an important factor affecting the predictive performance of the model. However, in this study, the choice of lag order still depends on the characteristics of the data and model assumptions. Although we optimized this parameter through cross validation, the selection of lag order may still need further adjustment in different datasets and application scenarios. Therefore, future research can explore more adaptive feature selection methods to better adapt to different patient populations and disease characteristics.

Future research can also combine deep reinforcement learning (DRL) technology to achieve more refined dynamic management of individual health data. Deep reinforcement learning can automatically adjust strategies based on environmental changes and provide personalized intervention measures. By adjusting health intervention strategies in real-time, DRL models are expected to achieve more significant results in stroke risk management. In addition, the combination of high-dimensional health data such as genomics, metabolomics, and environmental factors in modern medicine can provide more comprehensive information for stroke risk assessment. Future research can further improve the accuracy of prediction models through multi-source data fusion.

**Limitation**

Although this study achieved positive results, several limitations remain. First, the choice of lagged health indicators is dependent on the data characteristics and model assumptions. While optimization was performed through cross-validation, the choice of lag order may still need further adjustment in different datasets and application contexts. Second, although machine learning models can handle high-dimensional data, there may still be issues with overfitting or limited generalization ability. Additionally, the CHARLS dataset used in this study is primarily from China, and its applicability to other regions or cultural contexts may require further validation.

Moreover, while dynamic causal inference features were introduced, other important factors such as genomics, metabolomics, and high-dimensional health data were not fully considered, which could affect the model's comprehensiveness. Finally, although the model helps predict stroke risk, how to apply it in clinical practice and integrate it with physicians' diagnostic processes requires further research.

**Conclusion**

This study proposes a stroke risk prediction method that combines dynamic causal inference with machine learning models, significantly improving prediction accuracy and revealing key health factors that affect stroke. The research results indicate that dynamic causal inference features have important value in predicting stroke risk, especially in capturing the impact of changes in health status over time on stroke risk. By further optimizing the model and introducing more variables, this study provides theoretical basis and practical guidance for future stroke prevention and intervention strategies.

**Declarations**
- **Ethics approval and consent to participate**

This study protocol has been approved by the Ethics Review Committee of Peking University (Approval Numbers: Main Household Survey IRB00001052-11015, Biomarker Collection IRB00001052-11014). After a detailed explanation of the risks and benefits of participation, all participants signed informed consent forms allowing the sharing of their data and agreeing to store the data in the designated repository. The study adheres to the principles outlined in the Declaration of Helsinki.
- **Consent for publication**

Written informed consent for publication was obtained from all participants.
- **Availability of data and materials**

All data that support the findings of the current study are available from the corresponding authors upon reasonable request. The CHARLS dataset is available at [https://charls.pku.edu.cn/].
- **Competing interests**

The authors declare no conflicting financial interest.
- **Funding**

There is no funding.

- **Acknowledgements**

Not applicable